\documentclass{article}
\usepackage{spconf, graphicx}
\usepackage{amsmath, amsfonts, amssymb}
\usepackage{booktabs} %
\usepackage{tabularx}
\usepackage{hyperref}
\usepackage{multirow} %
\usepackage{diagbox} %
\usepackage{color, url}
\usepackage{cite}   %
\usepackage{enumitem}
\setlist[description]{leftmargin=\parindent,labelindent=\parindent}

\newcolumntype{Y}{>{\centering\arraybackslash}X} %
\newcommand{\around}{{\raise.17ex\hbox{$\scriptstyle\sim$}}}
\newcommand{\hermconj}{\mathsf{H}}

\renewenvironment{thebibliography}[1]{
  \begin{oldthebibliography}{#1}
  \setlength{\itemsep}{0.em}
  \setlength{\parskip}{0.em}
}
{
  \end{oldthebibliography}
}

\makeatletter
\renewcommand\section{\@startsection {section}{1}{\z@}%
                                   {-2.5ex \@plus -1ex \@minus -.2ex}%
                                   {1.75ex \@plus.2ex}%
                                   {\normalfont\Large\bfseries}}
\renewcommand\subsection{\@startsection{subsection}{2}{\z@}%
                                     {-1.5ex\@plus -1ex \@minus -.2ex}%
                                     {1.ex \@plus .2ex}%
                                     {\normalfont\large\bfseries}}
\makeatother

\usepackage{caption} 
\title{End-to-end dereverberation, beamforming, and speech recognition with improved numerical stability and advanced frontend}

 \name{\it Wangyou~Zhang$^{1}$, Christoph~Boeddeker$^{2}$, Shinji~Watanabe$^{3}$, Tomohiro~Nakatani$^{4}$,\\\it Marc~Delcroix$^{4}$, Keisuke~Kinoshita$^{4}$, Tsubasa~Ochiai$^{4}$, Naoyuki~Kamo$^{4}$,\\\it Reinhold~Haeb-Umbach$^{2}$, Yanmin~Qian$^{1}$
 }
 \address{
     $^1$MoE Key Lab of Artificial Intelligence, AI Institute, SpeechLab, Shanghai Jiao Tong University, China\\
     $^2$Paderborn University, Germany \hspace{1.5em}
     $^3$Johns Hopkins University, USA \hspace{1.5em}
     $^4$NTT Corporation, Japan
 }
\begin{document}
\ninept
\maketitle
\begin{abstract}
Recently, the end-to-end approach has been successfully applied to multi-speaker speech separation and recognition in both single-channel and multichannel conditions.
However, severe performance degradation is still observed in the reverberant and noisy scenarios, and there is still a large performance gap between anechoic and reverberant conditions.
In this work, we focus on the multichannel multi-speaker reverberant condition, and propose to extend our previous framework for end-to-end dereverberation, beamforming, and speech recognition with improved numerical stability and advanced frontend subnetworks including voice activity detection like masks.
The techniques significantly stabilize the end-to-end training process.
The experiments on the spatialized wsj1-2mix corpus show that the proposed system achieves about 35\% WER relative reduction compared to our conventional multi-channel E2E ASR system, and also obtains decent speech dereverberation and separation performance (SDR = 12.5 dB) in the reverberant multi-speaker condition while trained only with the ASR criterion.

\end{abstract}
\section{Introduction}
\label{sec:intro}
With the development of deep learning, much progress has been achieved in the speech processing field, including both speech enhancement \cite{Divide-Liu2019, Dual-Luo2020, Wavesplit-Zeghidour2020} in the frontend and automatic speech recognition (ASR) \cite{Comparative-Karita2019,Single-Tuske2020,Improved-Park2020} in the backend.
In recent years, more and more interests have been focused on the deep learning based speech processing in the cocktail party scenario \cite{Experiment-Cherry1953, Past-Qian2018}.
In this scenario, there are usually multiple speakers talking simultaneously, even with the presence of background noise and reverberation.
It is much more difficult to cope with than in the clean and anechoic conditions, and the ASR performance is still far behind humans in such conditions.

In the cocktail party scenario, while it is straightforward to combine separately trained speech enhancement and speech recognition components as one system, as investigated in many prior studies \cite{BLSTM-Heymann2015, Single-Isik2016}, the end-to-end (E2E) optimization of all involved components is also an important and interesting research topic.
The E2E system can naturally reduce the mismatch between different components through joint training.
In addition, only the noisy signal and the corresponding transcriptions are required for the E2E training of both frontend and backend, making it much easier for data collection and model training in real applications.
Some prior work has illustrated the potential of E2E optimized systems.
Settle et al. \cite{End-Settle2018} proposed a joint training framework, combining the chimera++ network \cite{Alternative-Wang2018} and end-to-end ASR \cite{Joint-Kim2017} for single-channel multi-speaker speech separation and recognition.
In the multichannel condition, the neural beamformer \cite{Neural-Heymann2016, Improved-Erdogan2016} based speech enhancement is often applied to better utilize the spatial information.
\emph{(1) Single-speaker cases:}
In \cite{Deep-Xiao2016, Unified-Ochiai2017, Beamnet-Heymann2017}, the neural beamformer is jointly trained with the acoustic\,/\,end-to-end ASR model for denoising and speech recognition.
Subramanian et al. \cite{Investigation-Subramanian2019} further included dereverberation in the joint training, which is based on the weighted prediction error (WPE) \cite{Speech-Nakatani2010} algorithm.
\emph{(2) Multi-speaker cases:}
Chang et al. \cite{MIMO-Chang2019} proposed the MIMO-Speech architecture, where the beamformer is jointly trained with ASR to perform speech separation.

In this paper, we aim to build a robust framework for the fully end-to-end optimization of dereverberation, beamforming (denoising and separation), and speech recognition.
In our prior work \cite{Endtoend-Zhang2020}, some preliminary attempts have been made to explore the end-to-end training of three components: WPE-based dereverberation, neural beamforming, and end-to-end ASR.
However, the well-known numerical instability issue \cite{Numerical-Lim2017} in operations of both WPE and beamforming, usually caused by the singularity in the matrix inverse operation, is still unsolved in \cite{Endtoend-Zhang2020}, leading to performance degradation or even misleading the model convergence.

In this work, we try to tackle this problem, by proposing four techniques to improve the stability and performance of the end-to-end system.
These methods have been proven extremely helpful in our setup, significantly mitigating the numerical instability issue during training.
Based on these techniques, we propose a robust architecture that supports the end-to-end training of different beamformer variants and ASR, which are also compared in our experiments.
In addition, the voice activity detection (VAD) like mask \cite{Iterative-Tu2019, Investigation-Subramanian2019} for WPE and beamforming is introduced to mitigate the frequency permutation problem in the end-to-end training, as described in Section~\ref{ssec:vad_mask}.
Our experiments on the spatialized wsj1-2mix \cite{MIMO-Chang2019} corpus show that the proposed approaches can achieve significant performance improvement compared to the previous system.

\vspace{-.75em}
\section{End-to-end framework for dereverberation, beamforming, and ASR}
\label{sec:framework}
\vspace{-.5em}
In this section, we first describe the proposed architecture for end-to-end dereverberation, beamforming (denoising and separation), and ASR.
And the formulation of different beamformer variants supported in the proposed framework is given.
We then introduce the techniques applied to solve the numerical instability issue.
Later, the frequency permutation phenomenon and our solution are discussed.
\begin{figure}[t]
  \centering
  \includegraphics[width=\columnwidth]{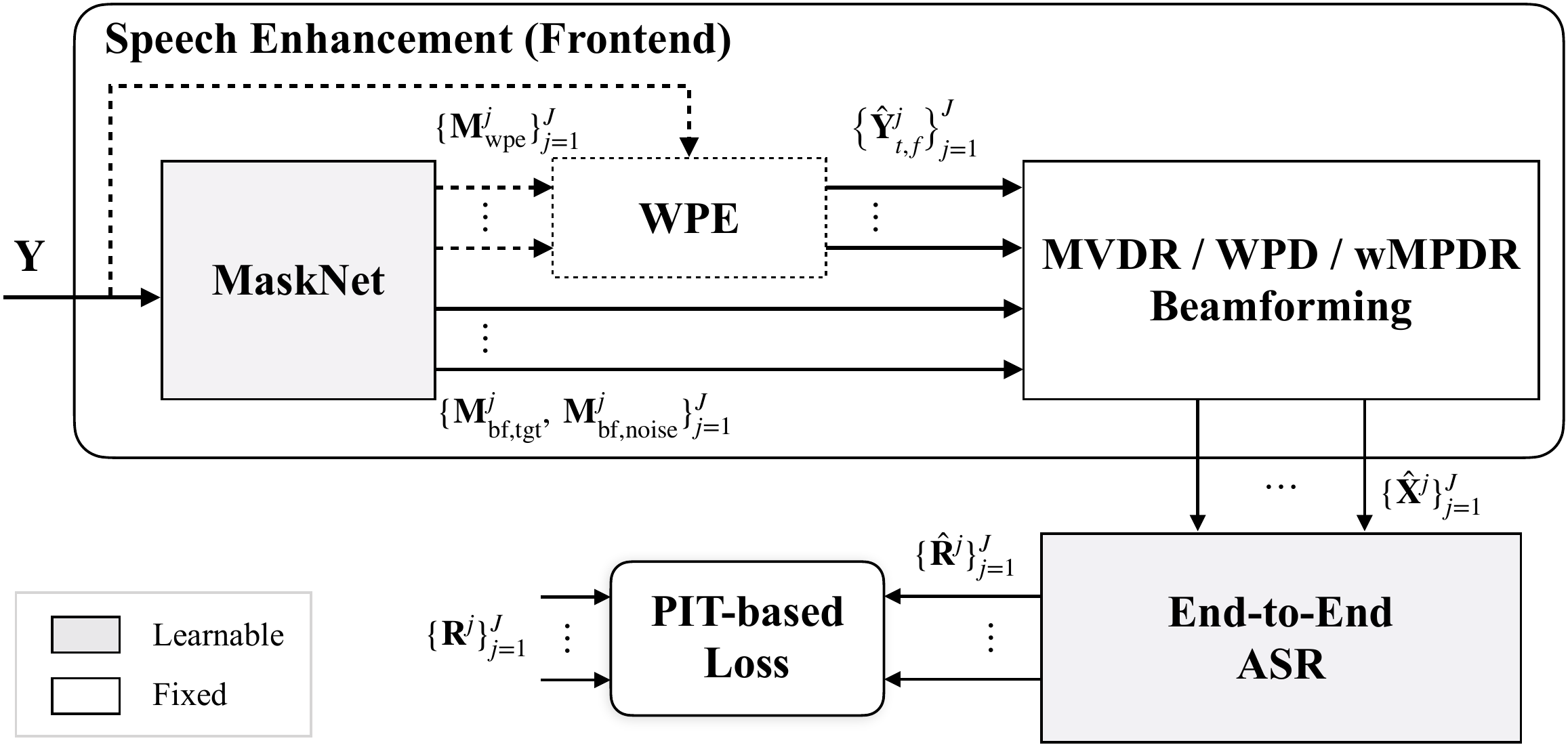}
  \caption{Proposed new architecture for end-to-end training of the frontend and ASR backend.}
  \label{fig:arch2}
\end{figure}
\vspace{-1.5em}
\subsection{Model architecture with advanced frontend}
\label{ssec:arch}
Our proposed end-to-end architecture is shown in Fig.~\ref{fig:arch2}, which is comprised of two main modules: the frontend (speech enhancement) and the backend (ASR).
Here, speech enhancement includes dereverberation, denoising and source separation.
In our previous article \cite{Endtoend-Zhang2020}, we adopted a weighted power minimization distortionless response (WPD) convolutional beamformer \cite{Unified-Nakatani2019} as a unified frontend, while the recent study \cite{Jointly-Boeddeker2020} showed that a WPD can be factorized into a WPE dereverberation filter and a weighted minimum power distortionless response (wMPDR) \cite{Jointly-Boeddeker2020} beamformer without loss of optimality when they are jointly optimized.
Therefore, in this article, we adopt the factorized form as a simpler alternative\footnote{The experimental result on WPD is also given in Table~\ref{tab:bf} for comparison.}.
That is, the frontend is composed of a single mask estimator (\texttt{MaskNet}), a DNN-WPE \cite{Neural-Kinoshita2017} dereverberation module, and a beamformer module.
In addition, we mainly support two alternative beamformer types, respectively, based on 1) minimum variance distortionless response (MVDR) \cite{Beamforming-Van1988} and 2) wMPDR.
While MVDR is a widely used state-of-the-art beamformer, wMPDR is shown to perform optimal processing jointly with WPE \cite{Jointly-Boeddeker2020}.
The ASR backend is a joint connectionist temporal classification (CTC)\,/\,attention-based encoder-decoder \cite{Joint-Kim2017} model for recognizing the separated single-channel speech.
Compared to those in our previous work \cite{Endtoend-Zhang2020}, the proposed architecture can support different beamformer variants in a single framework, by using a single mask estimator for WPE\,/\,beamforming and applying single-source WPE for processing speech of different sources.

Below we give a detailed description of the proposed system.
Consider a multichannel input speech signal composed of $J$ speakers, $\mathbf{Y}_{t,f} = \{Y_{t,f,c}\}_{c=1}^C \in \mathbb{C}^{C}$, it can be described as follows in the short-time Fourier transform (STFT) domain:
{%
\setlength{\abovedisplayskip}{2pt}
\setlength{\belowdisplayskip}{2pt}
\begin{align}
  \mathbf{Y}_{t,f} &= \sum_{j=1}^J \mathbf{X}^j_{t,f} + \mathbf{N}_{t,f} = \sum_{j=1}^J \mathbf{X}^{\text{(d)},j}_{t,f} + \mathbf{X}^{\text{(r)},j}_{t,f} + \mathbf{N}_{t,f} \label{eq:signal} \\
  \mathbf{X}^{\text{(d)},j}_{t,f} &= \sum_{\tau=0}^{\Delta-1} \mathbf{a}_{\tau,f}^j s_{t-\tau,f}^j \approx \mathbf{v}_f^j s_{t,f}^j \,, \label{eq:early_signal} \\
  \mathbf{X}^{\text{(r)},j}_{t,f} &= \sum_{\tau=\Delta}^{L_a} \mathbf{a}_{\tau,f}^j s_{t-\tau,f}^j \,, \label{eq:reverb_signal}
\end{align}
}
where $C > 1$ denotes the number of microphones.
$t \in \{1,\dots,T\}$ and $f \in \{1,\dots,F\}$ represent the indices of time and frequency bins.
$\mathbf{N}$ denotes noise.
$\mathbf{X}^j$ denotes the reverberant signal, which can be decomposed into an ``early'' part $\mathbf{X}^{\text{(d)},j}$ and a ``late'' part $\mathbf{X}^{\text{(r)},j}$.
$\mathbf{X}^{\text{(d)},j}$ contains the direct path and early reflection of the $j$-th speaker, while $\mathbf{X}^{\text{(r)},j}$ denotes the late reverberation.
$\mathbf{a}_{\tau,f}^j$ is the acoustic transfer function with length $L_a$.
$\Delta$ denotes the starting frame for the ``late'' part.
$s^j$ is the $j$-th source signal.
$\mathbf{v}^j_f = \big\{v_{f,c}^{j}\big\}_{c=1}^C \in \mathbb{C}^{C}$ is the steering vector (SV).
The input signal is first processed by the frontend module for dereverberation and separation.
First, the WPE submodule performs dereverberation separately for each source $j$ directly on the mixture $\mathbf{Y}$ in Eq.(\ref{eq:signal}):
{\allowdisplaybreaks
\begin{align}
  &\{\mathbf{M}_{\text{wpe}}^j\}_{j=1}^J,\ \{\mathbf{M}_{\text{bf},\text{tgt}}^j\}_{j=1}^J,\ \{\mathbf{M}_{\text{bf},\text{noise}}^j\}_{j=1}^J = \operatorname{MaskNet}(\mathbf{Y}) \,, \label{eq:mask_bf} \\
  &\lambda_{t,f}^j = \frac{1}{C}\sum_{c=1}^C \frac{M_{\text{wpe},t,f,c}^j}{\frac{1}{T}\sum_{\tau=1}^T M_{\text{wpe},\tau,f,c}^j} \left|Y_{t,f,c}\right|^2 \ \,\,\in \mathbb{R} \label{eq:power} \,, \\
  &\hat{\mathbf{Y}}^j = \mathbf{Y}^j_{\text{wpe}} = \operatorname{WPE}(\mathbf{Y}, \boldsymbol{\lambda}^j) \qquad\qquad\qquad\enspace\in \mathbb{C}^{T \times F \times C}\,. \label{eq:wpe}
\end{align}
}Here, $\mathbf{M}_{\text{wpe}}^j = \{M^j_{\text{wpe},t,f,c}\}_{t,f,c}$ denotes the estimated dereverberation mask, $\mathbf{M}^j_{\text{bf},\text{tgt}}$ and $\mathbf{M}^j_{\text{bf},\text{noise}}$ denote the estimated speech mask and distortion mask for the $j$-th speaker, respectively.
$\boldsymbol{\lambda}^j = \{\lambda_{t,f}\}_{t,f}^j$ is the estimated time-varying power of the speech signal.
$\operatorname{WPE}(\cdot)$ represents the dereverberation filter computation based on the WPE algorithm described in \cite{Nara-Drude2018}, and the detailed formulas are omitted here for simplicity.
The signal $\hat{\mathbf{Y}}$ is then denoised and separated by the neural beamformer. Within the scope of this paper, although different beamformers are designed for different objectives with a linear constraint, their solutions can be uniformly written as:
{\allowdisplaybreaks
\setlength{\abovedisplayskip}{2pt}
\setlength{\belowdisplayskip}{3pt}
\begin{align}
  \mathbf{\Phi}^{j}_{\alpha, f} &= \frac{\sum_{t=1}^T \overline{M}_{t,f}^{j} \hat{\mathbf{Y}}_{t,f}^j \big(\hat{\mathbf{Y}}_{t,f}^{j}\big)^{\hermconj}}{\sum_{t=1}^T \overline{M}_{t,f}^{j}} \hspace{2.8em}\in \mathbb{C}^{C \times C} \,, \label{eq:psd} \\
  \mathbf{w}^j_f &= 
    \frac{\big(\mathbf{\Phi}^{j}_{\text{N},f}\big)^{-1} \mathbf{\Phi}^{j}_{\text{S},f}}{\text{Trace}\left[\big(\mathbf{\Phi}^{j}_{\text{N},f}\big)^{-1} \mathbf{\Phi}^{j}_{\text{S},f}\right]} \mathbf{u} \,, \qquad\enspace\, \text{[w/o SV]} \label{eq:bf_filter} \\
    & = \frac{\big(\mathbf{\Phi}^{j}_{\text{N},f}\big)^{-1} \mathbf{v}^{j}_{f}}{\big(\mathbf{v}^{j}_{f}\big)^{\hermconj} \big(\mathbf{\Phi}^{j}_{\text{N},f}\big)^{-1} \mathbf{v}^{j}_{f}} \Big(v_{f,q}^{j}\Big)^* \,, \quad\enspace \text{[w/ SV]} \label{eq:bf_filter2} \\
  \hat{X}^j_{t,f} &= \big(\mathbf{w}^j_f\big)^{\hermconj} \hat{\mathbf{Y}}_{t,f} \hspace{8.6em}\in \mathbb{C} \,, \label{eq:beambformed}
\end{align}
}where $\overline{M}^j_{t,f} = \frac{1}{C}\sum_{c=1}^{C} M^j_{t,f,c}$ is a channel-averaged mask, where $M^j_{t,f,c} \in \{M^j_{\text{bf},\text{tgt},t,f}, M^j_{\text{bf},\text{noise},t,f}\}$, and $\mathbf{\Phi}_{\alpha,f}^j$ is a covariance matrix with a subscript $\alpha \in \{\text{N}, \text{S}, \text{noise}\}$.
We set $\overline{M}^j_{t,f} = \overline{M}_{\text{bf},\text{noise},t,f}^{j}$ for $\mathbf{\Phi}_{\text{noise},f}^j$ and $\overline{M}^j_{t,f} = \overline{M}_{\text{bf},\text{tgt},t,f}^{j}$ for $\mathbf{\Phi}_{\text{S},f}^j$. Similarly, we set $\overline{M}^j_{t,f} = \overline{M}_{\text{bf},\text{noise},t,f}^{j}$ and $\overline{M}^j_{t,f} = 1 / \lambda_{t,f}^j$, respectively, for $\mathbf{\Phi}_{\text{N},f}^j$ of MVDR and wMPDR.
$(\cdot)^*$ and $(\cdot)^{\hermconj}$ denote conjugate and conjugate transpose, respectively.
$\mathbf{w}^j_f$ is the beamforming filter for the $j$-th speaker, which can be calculated with either Eq.~(\ref{eq:bf_filter}) [w/o SV] or Eq.~(\ref{eq:bf_filter2}) [w/ SV].
While Eq.~(\ref{eq:bf_filter}) has been widely used for the E2E training of neural beamformers \cite{Unified-Ochiai2017, MIMO-Chang2019}, Eq.~(\ref{eq:bf_filter2}) is a standard equation for distortionless beamformers.
$\mathbf{u}$ is a vector denoting the reference channel, which can be estimated by the attention mechanism \cite{Multichannel-Ochiai2017}, or based on the average estimated a posteriori SNR \cite{Improved-Erdogan2016}, or manually set as a one-hot vector.
The subscript $q$ denotes the reference channel index.
$\hat{X}^j_{t,f}$ is the beamformed signal.
$\mathbf{v}_{f}^j$ can be calculated through the eigendecomposition \cite{Probabilistic-Ito2017}:
\vspace{-.75em}
\begin{align}
  &\mathbf{v}_{f}^j = \mathbf{\Phi}^{j}_{\text{noise},f} \operatorname{MaxEigVec}\left[\big(\mathbf{\Phi}^{j}_{\text{noise},f}\big)^{-1} \mathbf{\Phi}^{j}_{\text{S},f}\right] &\in&\ \mathbb{C}^{C} \,, \label{eq:steer_vec}
\end{align}
where $\operatorname{MaxEigVec}[\cdot]$ calculates the eigenvector corresponding to the maximum eigenvalue.
Due to the lack of complex eigendecomposition support in PyTorch at the time of writing, we replace it with the power iteration method \cite{Praktische-Mises1929}, which can be easily implemented for back-propagation, with a slight loss of precision.

It is worth noting that in the sense of end-to-end training, the MVDR and wMPDR beamformers are potentially equivalent.
By substituting the $\mathbf{\Phi}_{\text{N},f}^j$ for wMPDR defined above into Eq.~(\ref{eq:bf_filter}) or Eq.~(\ref{eq:bf_filter2}), we can find that the average operation in the denominator of Eq.~(\ref{eq:power}) is canceled.
Thus the derived wMPDR filter only depends~on the (inversed) mask predicted by the neural network, which is very similar to the MVDR formulation.
So it is hard to tell which beamformer is actually learned by the network via end-to-end training.

Finally, the separated stream $\hat{\mathbf{X}}^j = \{\hat{X}_{t,f}^j\}_{t,f}$ of each speaker $j$ is fed into the ASR backend for recognition.
First, the log Mel-filterbank coefficients $\mathbf{O}^{j}=\{\mathbf{o}^{j}_1,\dots,\mathbf{o}^{j}_T\}$ with global mean and variance normalization ($\text{GMVN-LMF}(\cdot)$) is extracted from $\hat{\mathbf{X}}^j$, which is then transformed by the encoder into a high-level representation $\mathbf{H}^{j}=\{\mathbf{h}^{j}_1,\dots,\mathbf{h}^{j}_L\} \ (L \leq T)$ with subsampling.
In order to solve the label ambiguity problem with multiple speakers ($J > 1$), the permutation invariant training (PIT) technique \cite{Permutation-Yu2017} is applied in the CTC module to determine the order of the label sequences.
With the best permutation derived in CTC, the representation $\mathbf{H}^{j}$ is processed by the attention-based decoder to generate the output token sequences $\hat{\mathbf{R}}^{j}=\{\hat{R}^{j}_1,\dots,\hat{R}^{j}_N\}$ with length $N$,
while $\mathbf{R}^j$ in Fig.~\ref{fig:arch2} is the corresponding reference label.
The speech recognition process for each speaker $j$ is formulated as follows:
{%
\setlength{\abovedisplayskip}{2pt}
\setlength{\belowdisplayskip}{2pt}
\begin{align}
  \mathbf{O}^{j} &= \operatorname{GMVN-LMF}(|\hat{\mathbf{X}}^{j}|) \,, \\[-1ex]
  \mathbf{H}^{j} &= \operatorname{Encoder}(\mathbf{O}^j) \,, \\[-1ex]
  \hat{R}^{j}_n &\sim \operatorname{Attention-Decoder} (\mathbf{H}^{j}, \hat{R}^{j}_{n-1}) \,,
\end{align}
}where $\hat{R}^j_n$ is the output token at the $n$-th decoding step.

Note that the entire system is optimized with sorely the ASR loss, which is a combination of the attention and CTC losses.

\vspace{-.3em}
\subsection{Attacking the numerical instability issue}
\label{ssec:numerical}
The numerical instability issue has been a well-known problem in the beamformer \cite{Numerical-Chakrabarty2015}, especially when optimized in an end-to-end manner.
The numerical problem generally originates from the complex operations in the WPE and beamforming formulas, such as the complex matrix inverse, leading to poor performance in certain frequency bins sparsely populated.
Such behaviors are particularly undesirable in the joint training with ASR, as they can easily result in not-a-number (NaN) gradients that fail to backpropagate correctly and even prevent the model from converging properly \cite{Endtoend-Zhang2020}, thus badly impacting the overall model performance.
In order to mitigate this problem, we propose four approaches to improve the stability of both WPE and beamforming submodules:
\vspace{-.3em}
\begin{description}[style=unboxed,leftmargin=0cm]
\itemsep 0em
\item[(1) Diagonal loading]
    In order to stabilize the matrix inverse operation in WPE and beamforming in Eqs.~(\ref{eq:wpe}), (\ref{eq:bf_filter}), (\ref{eq:bf_filter2}) and (\ref{eq:steer_vec}), particularly at its backward pass, we introduce a diagonal loading \cite{Numerical-Chakrabarty2015} term as a perturbation to the complex matrix $\mathbf{\Phi}$ before inversion:
    {%
    \setlength{\abovedisplayskip}{2.5pt}
    \setlength{\belowdisplayskip}{2.5pt}
    \begin{align}
      \mathbf{\Phi}^{\prime} &= \mathbf{\Phi} + \varepsilon \operatorname{Trace}(\mathbf{\Phi}) \mathbf{I} \,, \label{eq:diag}
    \end{align}
    }where $\mathbf{I}$ is the identity matrix, and $\varepsilon$ is a tiny constant.
    For better stabilization, $\operatorname{Trace}(\mathbf{\Phi})$ is used to make the term adaptive to signal level, and $\varepsilon$ was set at a relatively large value for WPE in our experiments, as described in Section~\ref{ssec:exp_setup}.
\item[(2) Mask flooring]
    When optimizing masks with an implicit criterion, i.e. the ASR loss, we observed that the mask estimator learned to predict sparse or spiky masks.
    This means, the mask estimator sets only the most relevant time-frequency bins to one, and the remaining ones to zero.
    It can then result in a singular covariance matrix in some frequency bins, making the WPE\,/\,beamforming process unstable.
    To avoid the spiky masks, we propose a mask flooring operation to introduce some regularization to the masks in Eq.~(\ref{eq:mask_bf}):
    {%
    \setlength{\abovedisplayskip}{2.5pt}
    \setlength{\belowdisplayskip}{2.5pt}
    \begin{align}
      \hat{M}_{t,f} &= \operatorname{Maximum}\{ M_{t,f},\ \xi \} \,, \label{eq:flooring}
    \end{align}
    }where $\hat{M}_{t,f}$ denotes the floored mask value, $M_{t,f} \in \{M_{\text{wpe},t,f,c},$ $\overline{M}_{\text{bf},\text{tgt},t,f},$ $\overline{M}_{\text{bf},\text{noise},t,f} \}$, and $\xi$ is a constant flooring factor.
    The idea of the flooring is, that enough values have to be nonzero to reduce the effect of the flooring value.
    So the mask estimator is prevented from predicting sparse or spiky masks.

\item[(3) More stable complex matrix operations]
    Due to the lack of complex support in PyTorch, the alternative method in Section 4.3 in \cite{Matrix-Petersen2012} was used in our previous work \cite{Endtoend-Zhang2020}, which tries to find a factor to construct an invertible real matrix and maps the complex inversion to some real matrix operations.
    But it sometimes fails due to the poor estimate of the factor that results in a singular matrix.
    In this paper, a more stable matrix inverse formula \cite{Note-Smith1974} is implemented, which converts the problem of complex matrix inverse $\mathbf{\Phi}^{-1} = (\mathbf{A} + i\mathbf{B})^{-1} \in \mathbb{C}^{m \times m}$ into the inverse of a $2m \times 2m$ real matrix:
    {%
    \setlength{\abovedisplayskip}{2pt}
    \setlength{\belowdisplayskip}{4pt}
    \begin{align}
      \begin{bmatrix}
          \mathbf{A} & \mathbf{B} \\
          -\mathbf{B} & \mathbf{A}
      \end{bmatrix}^{-1} &=
      \begin{bmatrix}
          \mathcal{R}\{\mathbf{\Phi}^{-1}\} & \mathcal{I}\{\mathbf{\Phi}^{-1}\} \\
          -\mathcal{I}\{\mathbf{\Phi}^{-1}\} & \mathcal{R}\{\mathbf{\Phi}^{-1}\}
      \end{bmatrix} \,, \label{eq:new_inverse}
    \end{align}
    }where $\mathcal{R}\{\cdot\}$ and $\mathcal{I}\{\cdot\}$ denotes the real and imaginary parts of a complex matrix.
    Furthermore, we replace the inverse and the subsequent multiplication operations in Eqs.~(\ref{eq:wpe}), (\ref{eq:bf_filter}) and (\ref{eq:bf_filter2}) with a solve operation, which directly computes the solution $\mathbf{x}$ to a linear matrix equation $\mathbf{\Phi}\mathbf{x} = \mathbf{v}$, where $\mathbf{x}$ and $\mathbf{v}$ are $m$-dimensional vectors.
    It further improves the numerical accuracy and stability.\footnote{The new implementations \texttt{inverse2} and \texttt{solve} are now available at \url{https://github.com/kamo-naoyuki/pytorch_complex}.}

    \item[(4) Double precision]
    In terms of the implementation, while the end-to-end systems normally operate with the single-precision data\,/\,parameters, we find it beneficial to use the double precision for complex operations in the frontend module.
    It can reduce the error caused by complex operations, such as the inverse of close-to-singular matrices.
    Thus the stability of matrix inverse related operations can also be improved.
    Similar effects are also reported in \cite{Joint-Heymann2019}, which proposes to jointly optimize the WPE and acoustic models.
\end{description}

With the above proposed techniques, we are now able to optimize the convolutional beamformer and ASR jointly, without the need of pretraining as in \cite{Endtoend-Zhang2020}.

\vspace{-.3em}
\subsection{VAD-like mask for WPE and beamforming}
\label{ssec:vad_mask}
\vspace{-.1em}
During the end-to-end optimization of the frontend and backend, we often observed that beamformer outputs corresponding to different speakers are permuted with each other at certain frequencies.
This is known as the frequency permutation problem \cite{Beamforming-Ikram2002}.
It is probably caused by the fact that beamforming filters are estimated independently at each frequency bin with the predicted time-frequency (T-F) masks,
and that the log Mel-filterbank features used for evaluating the ASR loss are obtained by averaging frequency bins with a triangle window, thus largely reducing the influence of the permutation errors on the loss.
This, however, is not optimal for speech enhancement in the frontend.
To solve this problem, instead of using T-F masks in Eq.~(\ref{eq:mask_bf}), we propose to use the voice activity detection (VAD) like masks \cite{Iterative-Tu2019, Investigation-Subramanian2019}, which share the same (soft) value over the frequency axis.
This mask will be shown effective to mitigate the frequency permutation problem in our experiments..

\begin{table}[t]
    \caption{Evaluation of the proposed techniques with the WPE + MVDR + ASR model of different architectures on the spatialized reverberant wsj1-2mix evaluation set. The number of filter taps $K$ and channels $C$ are set to 5 and 2 for evaluation (same as training), respectively.\protect\footnotemark}
    \label{tab:tricks}
    \centering
    \resizebox{0.84\columnwidth}{!}{%
    \begin{tabular}{l|cccc}
        \toprule
        Architecture & WER (\%) & PESQ & STOI & SDR (dB) \\
        \midrule
        Original mixture & - & 1.20 & 0.65 & -1.45 \\
        Arch in \cite{Endtoend-Zhang2020} & 21.88 & 1.12 & 0.62 & 1.23 \\
        \hline
        $\ \ $+ (1) Diagonal loading & 15.51 & 1.32 & 0.74 & \textbf{3.20} \\
        $\ \ $+ (2) Mask flooring & 20.13 & 1.24 & 0.71  & 1.14 \\
        $\ \ $+ (3) Stable complex op. & 15.70 & 1.31 & 0.74 & 3.05 \\
        $\ \ $+ (4) Double precision & 18.06 & 1.27 & 0.73 & 1.99 \\
        $\ \ $+ Tech (1)--(4) & 15.18 & 1.31 & 0.74 & 2.85 \\
        \hline
        Proposed arch & \multirow{2}{*}{\textbf{15.01}} & \multirow{2}{*}{1.31} & \multirow{2}{*}{0.74} & \multirow{2}{*}{2.81} \\
        $\ \ $+ Tech (1)--(4) &  &  &  &  \\
        \bottomrule
    \end{tabular}%
    }
\end{table}
\vspace{-.9em}
\section{Experiments}
\label{sec:exp}
\vspace{-.5em}
\subsection{Experimental setup}
\label{ssec:exp_setup}
In this section, we evaluate our proposed framework on the artificially generated spatialized wsj1-2mix dataset \cite{MIMO-Chang2019}, which contains anechoic and reverberant versions of multichannel two-speaker speech mixtures.
We trained our models on a multi-condition training subset, including both reverberant and anechoic training samples in the spatialized wsj1-2mix (98.5 hr $\times 2$), and WSJ train\_si284 single-speaker clean data (81.5 hr, only for training ASR).
Since the proposed framework jointly optimizes the frontend and backend with the ASR loss, no parallel clean data is required for training.
The development and evaluation subsets only contain reverberant samples from the spatialized wsj1-2mix, with the duration of 1.3 hr and 0.8 hr respectively.
For feature extraction, the STFT is performed on the 16-kHz input speech with a 25-ms Hann window and a 10-ms frame shift, and the 257-dimensional spectral feature is extracted.
For the ASR backend, 80-dimensional log Mel-filterbank features are extracted for each separated spectrum.
\footnotetext{More detailed results can be found at \url{https://speechlab.sjtu.edu.cn/members/wangyou-zhang/icassp21-material.pdf}.}

All our proposed models are implemented based on the ESPnet framework.
The mask estimation network in Fig.~\ref{fig:arch2} is a 3-layer bidirectional long-short term memory (BLSTM) network with 600 cells in each direction, followed by $J \times 3$ output layers, where the number of speakers is $J = 2$.
The number of iterations for performing WPE is 1.
During training, the number of channels $C$ and WPE filter taps $K$ are fixed to 2 and 5, respectively.
In the ASR backend, we followed the same configurations in \cite{End-Chang2020}.
We set $\varepsilon$ in Eq.~(\ref{eq:diag}) to $10^{-3}$ and $10^{-8}$ for WPE and beamforming, respectively.
The mask flooring factor $\xi$ in Eq.~(\ref{eq:flooring}) is set to $10^{-6}$ and $10^{-2}$ for WPE and beamforming, respectively.
The number of iterations for estimating the steering vector using the power iteration is set to 2.
The reference channel $q$ in Eq.(\ref{eq:bf_filter}) is set to 1.
The Noam optimizer with 25000 warmup steps and an initial learning rate of 1.0 was used for training.

 \vspace{-.5em}
\subsection{Experimental results}
\label{ssec:result}
We first evaluate the proposed techniques in Section~\ref{ssec:numerical} in both previously used \cite{Endtoend-Zhang2020} and the proposed architectures, as shown in Table~\ref{tab:tricks}.
For speech recognition, we use the word error rate (WER) for evaluation.
The speech enhancement (SE) performance is evaluated using three common metrics: signal-to-distortion ratio (SDR) \cite{BSSEAVL-Fevotte2005}, short-time objective intelligibility (STOI) \cite{STOI-Jensen2016} and perceptual evaluation of speech quality score (PESQ) \cite{PESQ-Rix2001}.
And the clean source signal from WSJ is adopted as the reference signal.
In Table~\ref{tab:tricks}, we can observe that all proposed techniques can bring significant performance improvement compared to the baseline architecture in \cite{Endtoend-Zhang2020}.
And the combination of the four techniques can further achieve a better ASR result, with improved speech enhancement performance.
This illustrates the effectiveness of the proposed approaches.
The last row shows that with the proposed techniques, our proposed architecture in Fig.\ref{fig:arch2} can also achieve comparable performance.

We then evaluate the proposed architectures with different beamformer variants under different configurations of filter taps $K \in \{1,3,5,7,$ $10\}$, while the number of channels $C$ is fixed to 6, and only present the best performance of each model in Table~\ref{tab:bf} due to the limited space.
We also present the best ASR results from \cite{Endtoend-Zhang2020} in rows 2 and 3 for comparison, and the SE performance are also evaluated.
Comparing rows 2 \& 5 and rows 3 \& 6, we can observe that the proposed methods greatly improve the ASR and SE performances compared to the previous systems, which attributes to the proposed techniques for mitigating the numerical instability issue.
From row 4 to row 5, the performance gain indicates the DNN-WPE submodule plays an important role in our proposed architecture.
Comparing the second and third sections in Table~\ref{tab:bf}, the MVDR and wMPDR beamformers show very similar results based on the formulas in either Eq.~(\ref{eq:bf_filter}) [w/o SV] or Eq.~(\ref{eq:bf_filter2}) [w/ SV].
This also indicates the potential equivalence of these beamformers in the end-to-end training, as mentioned in Section~\ref{ssec:arch}.
And the latter formula tends to yield better ASR results with end-to-end training.
When comparing the second and the last sections in Table~\ref{tab:bf}, we can find that the proposed VAD-like masks are beneficial for the SE performance, with obvious improvement on PESQ, STOI and SDR.
This indicates that the VAD-like mask can effectively mitigate the frequency permutation problem, thus improving the SE performance.
Since the evaluation set is generated based on the WSJ eval92 subset, the first row in Table~\ref{tab:bf} can be regarded as the topline for our system.
And we can observe that the proposed models with different beamformer variants can all achieve very good ASR performance, with an only \around 5\% higher WER than the topline on WSJ.

\begin{table}[t]
    \caption{Evaluation of different beamformer variants and mask types on the spatialized reverberant wsj1-2mix evaluation set. ``w/ SV'' and ``w/o SV'' in Eq.~(\ref{eq:bf_filter})--(\ref{eq:bf_filter2}) denote with and without explicit use of the steering vector, respectively.}
    \label{tab:bf}
    \centering
    \resizebox{\columnwidth}{!}{%
    \begin{tabularx}{1.05\columnwidth}{cl *{2}{Y}|>{\centering\arraybackslash\hsize=0.9cm}X >{\centering\arraybackslash\hsize=0.9cm}X >{\centering\arraybackslash\hsize=0.9cm}X >{\centering\arraybackslash\hsize=0.9cm}X}
        \toprule
        ID & Model (+ASR) & Formula & Mask & WER & PESQ & STOI & SDR \\
        \midrule
        1 & WSJ eval92 \cite{Comparative-Karita2019} & - & - & 4.4 & - & - & - \\
        2 & WPE+MVDR \cite{Endtoend-Zhang2020} & \multirow{1}{*}{w/o SV} & T-F & 15.72 & 1.15 & 0.62 & 0.62 \\ %
        3 & WPD \cite{Endtoend-Zhang2020} & \multirow{1}{*}{w/o SV} & T-F & 13.97 & 1.33 & 0.68 & 0.38 \\ %
        \hline
        4 & MVDR & \multirow{3}{*}{w/o SV} & \multirow{3}{*}{T-F} & 11.66 & 1.46 & 0.80 & 6.48 \\ %
        5 & WPE+MVDR & & & 9.50 & 1.56 & 0.83 & 7.73 \\ %
        6 & WPE+wMPDR & & & 9.44 & 1.63 & 0.82 & 8.49 \\ %
        \hline
        7 & WPE+MVDR & \multirow{2}{*}{w/ SV} & \multirow{2}{*}{T-F} & \textbf{9.02} & 1.50 & 0.83 & 6.93 \\ %
        8 & WPE+wMPDR & & & 9.23 & 1.54 & 0.82 & 7.12 \\ %
        \hline
        9 & WPE+MVDR & \multirow{2}{*}{w/o SV} & \multirow{2}{*}{VAD} & 9.45 & 1.95 & 0.86 & \textbf{12.54} \\ %
        10 & WPE+wMPDR & & & 10.26 & 1.97 & 0.86 & 12.20 \\ %
        \bottomrule
    \end{tabularx}%
    }
\end{table}

\vspace{-.75em}
\section{Conclusions}
\label{sec:conclusions}
\vspace{-.5em}
In this paper, we propose a robust framework for end-to-end training of dereverberation, beamforming (denoising and separation), and speech recognition.
Four techniques are proposed to regularize and stabilize the WPE\,/\,beamforming process in the frontend module, which are shown to effectively improve the numerical stability.
Different beamformer variants and mask types are compared in our proposed framework.
Our experiments on the spatialized wsj1-2mix corpus show that the proposed end-to-end system can achieve fairly good ASR results, with also decent speech enhancement performance in the reverberant multi-speaker condition, while only optimized with the ASR criterion.
In our future work, we would like to investigate the end-to-end training in realistic and more challenging conditions.

\vspace{-.75em}
\section{Acknowledgement}
\label{sec:ack}
\vspace{-.5em}
Wangyou Zhang and Yanmin Qian were supported by the China NSFC projects (No. 62071288 and U1736202).
The work reported here was started at JSALT 2020 at JHU, with support from Microsoft, Amazon and Google.
Experiments were carried out on the PI supercomputers at Shanghai Jiao Tong University.

\bibliographystyle{IEEEtran}
\bibliography{refs}

\end{document}